\documentstyle[12pt]{article}
\begin{document}
\title{Quantum theory of non-relativistic particles
interacting with gravity}
\author {C. Anastopoulos \\ Theoretical Physics Group, The Blackett Lab. \\
Imperial College \\ E-mail : can@tp.ph.ic.ac.uk}
\date {October 1995}
\maketitle

\begin{abstract}
\par
We investigate the effects of the gravitational field on the
quantum dynamics of non-relativistic particles. We consider N
non-relativistic particles, interacting with the linearized
gravitational field. Using the Feynman - Vernon influence functional
technique, we trace out the graviton field, to obtain a master
equation for the system of particles to first order in $G$. The
effective interaction
between the particles, as well as the self-interaction is non-local
in time and in general non-markovian. We show that the gravitational
self-interaction cannot be held responsible for decoherence of
microscopic particles due to the fast vanishing of the diffusion
function. For macroscopic particles though, it leads to
diagonalization to the energy eigenstate basis, a desirable feature
in gravity induced collapse models. We
finally comment on possible applications.
\end{abstract}

\pagebreak

\renewcommand {\theequation}{\thesection.\arabic{equation}}
\let \ssection = \section
\renewcommand{\section}{\setcounter{equation}{0} \ssection}

\section{Introduction}

\par
There has been recently a considerable interest in the application of
the influence functional technique \cite{FeVe} in the study of non-equilibrium
systems in physics. Besides quantum Brownian motion
\cite{CaLe,HPZ,UnZu,GSI} for which the method  was initially developed, it has
been
applied to the modelling of particle-field interactions \cite{HuM},
radiation damping \cite{BaCa}, black-body radiation \cite{Ang} and
most recently to non-inertial detectors coupled to a scalar field
\cite {AHR}. It is one of the most powerful techniques to obtain
master equations, when the coarse-graining comes from the splitting of
degrees of freedom to system and environment.
\par
In this paper we apply the technique in another case : a system
of N non-relativistic particles coupled to linearized gravity. A
motivation for this is the possibility that
 gravity induces
decoherence on the particles' states. This is a suggestion made on
different contexts on fundamental irreversibility in quantum mechanics
\cite{Pen,GRW,Zeh}.  The weakness of the coupling suggests that the
probable decoherence time should be very large, but the particular
form of the coupling (quadratic to momentum) and the possibility of
persistent noise might give rise to observable consequences.
\par
In addition, the model we present here can be generalized in a
straightforward way to obtain a description of systems of quantum
mechanical detectors of gravitational waves.
\par
Our model consists of non-relativistic particles coupled to the
linearized gravitational field, which is assumed to be initially at
its vacuum state.We argue that  a factorizing  initial condition is,
in contrast to quantum Brownian motion, well suited for our
system. The modes of the graviton field are bounded in energy by an
ultraviolet cut-off $\Lambda$, which on  physical grounds  should be
much smaller than the Compton wavelength of the particles. In
addition, we assume
that the particles are almost stationary. Our analysis resembles, in a
way, the one of \cite{AHR}. Like them, we obtain correlation kernels
describing a non-local interaction between the particles. The
influence functional we construct is rather different from the ones
considered  in the
literature, due to the particular features of the gravitational coupling
which is quadratically coupled to momenta. The result of our analysis
is the non-markovian master equation (3.9) . For the case of a  single
particle, it is simplified significantly (4.1). We see that the
dissipation and diffusion are determined solely from the Hamiltonian
operator. We can interpret our results as  a continuous monitoring of
the energy of the particle by the gravitational environment.
\par
The diffusion function, which is responsible for decoherence
vanishes at long times and it turns out, that unless we consider
macroscopically massive bodies, the rate of gravitationally induced
decoherence is extremely small. This is a desirable result in
connection with the gravity-induced collapse models.

\section{The model}

\par
Consider N particles on a 3+1 dimensional spacetime, moving on  trajectories
$({\bf x}_n(\tau_n),t_n(\tau_n))$ parametrized by the proper times
$\tau_n$ so that $t_n(\tau_n)$ is a strictly increasing function of
$\tau_n$ \cite{AHR}.
We assume that the gravitational interaction is very weak and
therefore work in the linearized approximation. That is, the metric
is:
\begin{equation}
g_{\mu \nu} = \eta_{\mu \nu} + h_{\mu \nu}
\end{equation}
with $\eta_{\mu \nu}$ the Minkowski space metric.
\par
We take the non-relativistic limit for the particles, that is, we
assume that there exists a frame, with respect to which they are
almost stationary and therefore can write their trajectories as
$(a^i_{(n)} + x^i_{(n)}(t),t)$, having identified the global time
coordinate
 $t$ with the
proper-time of the particles. We assume that $\mid {\bf x}_{(n)}\mid$ is much
smaller than the distance between any two particles $d_{nm} =
|{\bf a}_{(n)} - {\bf a}_{(m)} |$ . This is a good approximation as long
as $d_{nm}$ is much larger
than the maximum wavelength of the graviton field that can be
excited.
Essentially, we consider the particles moving around some fixed
sites coordinatized by ${\bf a}_{(n)}$, so that their individual motion does
not significantly change their distances. In any case, this
approximation does not affect at all the discussion on the self
interaction of the particles through the gravitational field.
\par
We work in  in the transverse-traceless gauge for the linearized
gravitational field
($h_{0 \mu} = 0$, $h^{i j }_{ i},  = 0$, $h^{i}_{i}=0$). Under these
approximations ,
the total action of
the system  for evolution from global time $t = 0$ to $t = T$, reads
\begin{equation}
S_{tot} = S_{gr} + S_{par} + S_{int}
\end{equation}
where
\begin{equation}
S_{gr} = \frac{1}{4 \pi G} \int_0^T dt \int d^3x h_{\mu \nu,\rho}
h^{\mu \nu, \rho} =  \\ \nonumber
\frac{1}{4 \pi G} \int_0^T dt \int d^3x
(\dot{h}_{ij} \dot{h}^{ij} - h_{ij,k}h^{ij,k})
\end{equation}
\begin{equation}
S_{par}= \sum_n \int_0^T dt \frac{1}{2} \delta_{ij} \dot{x}^i_{(n)}
\dot{x}^j_{(n)}
\end{equation}
\begin{equation}
S_{int} = \sum_n  \int_0^T dt h_{ij} \dot{x}^i_{(n)}
\dot{x}^j_{(n)}
\end{equation}
Note that we have set  $\hbar = c  = m = 1$.
\par
We expand the graviton field in normal modes:
\begin {equation}
h_{ij}({\bf x},t) = \int \frac{d^3k}{(2 \pi)^3} \sum_r ( q_{1 {\bf k}}^{(r)}
cos
{\bf k x} + q_{2 {\bf k}}^{(r)} sin {\bf k x} ) A_{{\bf k}
ij}^{(r)}
\end{equation}

The polarization matrices $ A_{{\bf k}ij}^{(r)}$ ( $r = 1,2$) are
traceless and transverse and can be chosen to satisfy :
\begin{equation}
\ A_{{\bf k}i}^{(r)j} A_{{\bf k}j}^{(r') l } = \delta_{r r'} (
\delta^i_l - \frac{k^i k_l}{k^2})
\end{equation}
\begin{equation}
\sum_r A_{{\bf k}ij}^{(r)} A_{{\bf k}kl}^{(r)}= (\delta_{(ij} -
\frac{k_{(i} k_j}{{\bf k}^2}) (\delta_{k)l} - \frac{k_{k)}k_l}{{\bf k}^2})
\\ \nonumber
:= T^{ijkl}({\bf k})
 \end{equation}
The gravity part of the action therefore reads:
\begin{equation}
S_{gr} = \frac{1}{2 \pi G} \int_0^T\int \frac{d^3k}{(2 \pi)^3} \sum_r
[(\dot{q}_{1 {\bf
k}}^{(r)2} + {\bf k}^2 q_{1 {\bf k}}^{(r)2})+(\dot{q}_{2 {\bf
k}}^{(r)2} + {\bf k}^2 q_{2 {\bf k}}^{(r)2})]
\end{equation}
This is just the action for two massless scalar fields propagating on
Minkowski spacetime.
\par
We now write the coupling part of the action
\begin{equation}
S_{int} = \frac{1}{2} \int_0^T dt \int \frac{d^3k}{(2 \pi)^3} \sum_r
\sum_n (q_{1 {\bf k}}^{(r)} cos {\bf k a}_{(n)}+ q_{2 {\bf k}}^{(r)}
sin{\bf k a}_{(n)}) A_{{\bf k}ij}^{(r)} \dot{x}^i \dot{x}^j
\end{equation}
where within our approximations we ignored the ${\bf x}_{(n)}$ terms
in the trigonometric functions.
\par
By using the collective index $\alpha$ to include the ${\bf k}$,$r$
and the indexing of our oscillator by $1$ or $2$, we write :
\begin{equation}
S_{gr} + S_{int} = \int_0^T \sum_{\alpha}[\frac{1}{2 \pi
G}(\dot{q}_{\alpha}^2 + \omega_{\alpha}^2 q_{\alpha}^2) + q_{\alpha}
J_{\alpha}]
\end{equation}
where
\begin{equation}
J_{{\bf k} 1}^{(r)} = cos {\bf k a_{(n)}}A_{{\bf k}ij}^{(r)}\dot{x}^i
\dot{x}^j
\end{equation}
\begin{equation}
J_{{\bf k} 1}^{(r)} = sin {\bf k a_{(n)}}A_{{\bf k}ij}^{(r)}\dot{x}^i
\dot{x}^j
\end{equation}
and $ \omega_{{\bf k}} = |{\bf k}| $.
 This is just the action of a collection of forced harmonic
oscillators.Therefore the total action is that of a collection of N
non-relativistic free particles
interacting with a bath of harmonic oscillators, through couplings
depending quadratically on the velocity.
\par
The tracing out of the graviton modes can be done exactly since the
path integral is a gaussian with respect to them. We compute the
influence functional:
\begin{eqnarray}
{\cal F}[{\bf x}(t),{\bf x'}(t')] = \hspace{10cm} \\ \nonumber
\prod_{\alpha \beta}\int
dq^{\alpha}_f \int dq'^{\beta}_f \int dq^{\alpha}_0
\int dq^{\beta}_0 \delta(q_{\alpha} - q'_{\beta}
 \int Dq_{\alpha}(t) Dq'_{\beta}(t') \hspace{3cm} \\ \nonumber
\exp [ i S_{gr}[q_{\alpha}(t)] +  i S_{int}[q_{\alpha}(t), {\bf x}(t)]
- i S_{gr}[q'_{\alpha}(t')] -  i S_{int}[q'_{\alpha}(t'), {\bf
x'}(t')]]  \\ \nonumber
\rho_0(q^{\alpha}_0,{\bf x}_0,q'^{\beta}_0,{/bf x'}_0) \hspace{6cm}
\end{eqnarray}
where the integration is over the paths satisfying: $q^{\alpha}(0) =
q^{\alpha}_0$, $q^{\alpha}(T) = q^{\alpha}_f$, $q'^{\alpha}(0) =
q'^{\alpha}_0$, $q^{\alpha}(T) = q'^{\alpha}_f$. Here $\rho_0$ is the
density matrix of the total system.
 The path integrations can be carried out exactly, to obtain:
\begin{eqnarray}
{\cal F}[{\bf x}(t),{\bf x'}(t')] = \hspace{10cm}  \nonumber \\
{\cal N}(T) \exp [ - \sum_{\alpha}
\frac{i}{2 \omega_{\alpha}}\int_0^T ds \int_0^s ds'
(J_{\alpha}+J_{\alpha}')(s) \sin \omega_{\alpha}(s-s') (J_{\alpha}-
J_{\alpha}')(s') \nonumber  \\
-\sum_{\alpha}\frac{1}{2 \omega_{\alpha}}\int_0^T ds \int_0^s ds'
(J_{\alpha}-J_{\alpha}')(s) \cos \omega_{\alpha}(s-s') (J_{\alpha}-
J_{\alpha}')(s')] \hspace{2cm}
\end{eqnarray}
\par
In deriving this we have assumed that at $t=0$ the states of the
particles and of the graviton field were uncorrelated and that the
field were on its vacuum state, i.e.
\begin{equation}
\Psi[h_{ij}] = C  \exp[\sum_{\alpha} \frac{i}{2\omega_{\alpha}}
{\bf q}_{\alpha}^2]
\end{equation}
This initial condition is usually considered unphysical in
quantum Brownian motion models. We believe that it is actually a quite
good one for the case of gravity. Graviton modes are excited only by
non-stationary particles. Therefore, this initial condition reflects an
operation on the particles of a very fast acceleration just before
$t=0$.
\par
Substituting the expressions for the currents $J_{\alpha}$ into
the influence functional we get:
\begin{eqnarray}
{\cal F}[{\bf x},{\bf x'}] = N(T) \exp[ i \sum_{n,m}\int_0^T ds \int_0^s
ds' (\dot{x}^i_{(n)}\dot{x}^j_{(n)} +
\dot{x'}^i_{(n)}\dot{x'}^j_{(n)})(s) \\ \nonumber
 \gamma^{ijkl}_{(n)(m)}(s-s') (\dot{x}^k_{(m)}\dot{x}^l_{(m)} -
\dot{x'}^k_{(m)}\dot{x'}^l_{(m)})(s') \\ \nonumber
-\sum_{n,m}\int_0^T ds \int_0^s
ds' (\dot{x}^i_{(n)}\dot{x}^j_{(n)} -
\dot{x'}^i_{(n)}\dot{x'}^j_{(n)})(s) \\ \nonumber
 \eta^{ijkl}_{(n)(m)}(s-s') (\dot{x}^k_{(m)}\dot{x}^l_{(m)} -
\dot{x'}^k_{(m)}\dot{x'}^l_{(m)})(s')]
\end{eqnarray}
\par
The kernels $\gamma_{(n)(m)}$ and $\eta_{(n)(m)}$ are given by the
expressions:
\begin{equation}
\gamma^{ijkl}_{(n)(m)}(s) = \frac{G}{8 \pi^2} \int \frac{d^3k}{\mid
{\bf k} \mid} \sin|{\bf k}|s \cos {\bf
k}({\bf a_n}- {\bf a_m}) T^{ijkl}({\bf k})
\end{equation}
\begin{equation}
\eta^{ijkl}_{(n)(m)}(s) = \frac{G}{8 \pi^2} \int \frac{d^3k}{\mid
{\bf k} \mid} \cos|{\bf k}|s \cos {\bf
k}({\bf a_n}- {\bf a_m}) T^{ijkl}({\bf k)}
\end {equation}
\par
These are the dissipation and noise kernels, similar to the ones
derived in \cite{AHR} for the case of detectors minimally coupled to
a scalar field. For $n \neq m$ they describe the dissipation and
diffusion induced on the particle $n$ from the particle $m$ , while
for $n=m$ they contain the effects of the self-interaction of the
particle through its interaction with the gravitational field.
\par
In order to keep them finite, we have to restrict the integration
range to values of $|{\bf k}|$ smaller than a cut-off $\Lambda$. This
is natural, since we do not expect the non-relativistic particles to
excite graviton modes with arbitrarily high energy. In fact $\Lambda$
should be much smaller than the Compton wavelength of the
particle. This is in accordance with our previous
approximations, since the distance between any particles remains much
larger than their Compton wavelength.
\par
In the particular case $n = m$ we can perform the angular integrations
in spherical coordinates in the equations for the kernels and obtain:
\begin{equation}
\gamma^{ijkl}_{(n)(n)}(s) =\frac{G}{15 \pi}
\delta^{ijkl}\int_0^{\Lambda} dk k \sin ks
\end{equation}
\begin{equation}
\eta^{ijkl}_{(n)(n)}(s) =\frac{G}{15 \pi}
\delta^{ijkl}\int_0^{\Lambda} dk k \cos ks
\end{equation}
We note that by taking the cut-off to infinity, the dissipation kernel
becomes essentially the derivative of a delta-function, as in the well
studied case of quantum Brownian motion with ohmic environment. The
corresponding semi-classical equations for $t >> \Lambda^{-1}$ can be
found using the standard
procedure \cite{HPZ,HuM,AHR}:
\begin{equation}
\ddot{x}^i + \frac{2G}{15} \delta_{ijkl} \ddot{x}^j \ddot{x}^k
\dot{x}^l = (\ddot{x}^l \delta^{ik} + \ddot{x}^k \delta^{il})
\xi_{kl}
\end{equation}
with $\xi_{kl}(t)$ a stochastic force determined by the correlator:
\begin{equation}
\langle \xi_{ij}(t) \xi_{kl}(t') \rangle = \eta^{ijkl}(t-t')
\end{equation}

\section{The master equation}

\par
Having obtained an expression for the influence functional we can
compute the reduced density matrix propagator:
\begin{equation}
J({\bf x_f},{\bf x'_f},t|{\bf x_0},{\bf x'_0},0) = \int \int D{\bf
x} D{\bf x'} \exp(i S_{par}[{\bf x}] - i S_{par}[{\bf x'}]) {\cal
F}[{\bf x},{\bf x'}]
\end{equation}
where the integration is over all paths ${\bf x}(s)$, ${\bf x'}(s')$
satisfying: ${\bf x}(0) = {\bf x_0}$, ${\bf x'}(0) = {\bf x'_0}$, ${\bf
x}(t) = {\bf x_f}$, ${\bf x'}(t) = {\bf x_f}$.
\par
The knowledge of the reduced density matrix propagator enables us
to construct a master equation. Our system is characterized from the
non-local dissipation and diffusion in the influence functional, and
the coupling which is quadratic to the velocities. Because of the
peculiarities of the latter, the general method of Hu,Paz and Zhang
\cite{HPZ} is not applicable here. Instead we compute the influence
functional perturbatively (first order in $G$) and use the Feynman
prescription for the determination of the master equation.
\par
Our starting point is the density matrix propagator for the free
particle under  external forces ${\bf F}(s)$, ${\bf F'}(s)$:
\begin{eqnarray}
J^{(0)}[{\bf F},{\bf F'}]({\bf x_f},{\bf x'_f},t|{\bf x_0},{\bf x'_0},0) =
\frac{C}{t}
\exp[ \frac{i}{2t}({\bf x_f} - {\bf x_0})^2 - \frac{i}{2t}({\bf x'_f}
- {\bf x'_0})^2 \\ \nonumber
+\frac{i}{t} {\bf x_0} \int_{0}^{t} ds s {\bf F}(s)-\frac{i}{t}
{\bf x'_0} \int_{0}^{t} ds s {\bf F'}(s) \\ \nonumber
+\frac{i}{t} {\bf x_f} \int_{0}^{t} ds (t-s) {\bf F}(s)-\frac{i}{t}
{\bf x'_f} \int_{0}^{t} ds (t-s) {\bf F'}(s) \\ \nonumber
+ \frac{i}{t} \int_{0}^t ds \int_{0}^{s} ds' s'(t-s){\bf F}(s) {\bf F}(s')
- \frac{i}{t} \int_{0}^t ds \int_{0}^{s} ds' s'(t-s){\bf F'}(s) {\bf F'}(s')]
\end{eqnarray}
The perturbation expansion of the propagator is writen then formally:
\begin{eqnarray}
J({\bf x_f},{\bf x'_f},t|{\bf x_0},{\bf x'_0},0) =
\\ \nonumber
{\cal F}[-i
\frac{\delta}{\delta {\bf F}(s)},i \frac{\delta}{\delta {\bf F'}(s)}]
J^{(0)}[{\bf F},{\bf F'}]({\bf x_f},{\bf x'_f},t|{\bf x_0},{\bf x'_0},0)
\mid_{{\bf F}={\bf F'}=0}
\end{eqnarray}
To first order in $G$ we obtain:
\begin{eqnarray}
J({\bf x_f},{\bf x'_f},t|{\bf x_0},{\bf x'_0},0) =
\frac{C}{t} \exp \sum_{(n)(m)} [ 4 G \delta_{ij} \delta_{kl}
g^{ijkl}_{(n)(m)}
\\ \nonumber
+ \frac{i}{2t} \delta_{ij} (x_f - x_0)^i_{(n)} (x_f - x_0)^j_{(n)} \delta_{mn}
- \frac{i}{2t} \delta_{ij} (x'_f - x'_0)^i_{(n)}  (x'_f - x'_0)^j_{(n)}
\delta_{nm}\\ \nonumber
-\frac{G}{t} (3f - 4 i g)^{ijkl}_{(n)(m)} \delta_{ij}(x_f - x_0)^i_{(n)} (x_f
- x_0)^j_{(n)} \delta_{nm} \\ \nonumber
-\frac{G}{t} (3f + 4 i g)^{ijkl}_{(n)(m)} \delta_{ij}(x'_f - x'_0)^i_{(n)} (x_f
- x_0)^j_{(n)} \delta_{nm} \\ \nonumber
- \frac{i G}{2 t^2} f^{ijkl}_{(n)(m)} [(x_f - x_0)^i_{(n)} (x_f -
x_0)^j_{(n)} (x_f
- x_0)^k_{(m)} (x_f - x_0)^l_{(m)}
\\ \nonumber
 - (x'_f - x'_0)^i_{(n)} (x'_f -
x'_0)^j_{(n)} (x'_f
- x'_0)^k_{(m)} (x'_f - x'_0)^l_{(m)} ]  \\ \nonumber
-\frac{G}{2 t^2} g^{ijkl}_{(n)(m)} [ (x_f - x_0)^i_{(n)} (x_f - x_0)^j_{(n)}
(x_f - x_0)^k_{(m)} (x_f - x_0)^l_{(m)} +
\\ \nonumber
  (x'_f - x'_0)^i_{(n)}
(x'_f - x'_0)^j_{(n)}
(x'_f - x'_0)^k_{(m)} (x'_f - x'_0)^l_{(m)}
\\ \nonumber
- 2 (x_f - x_0)^i_{(n)}
(x_f - x_0)^j_{(n)}
(x'_f - x'_0)^k_{(m)} (x'_f - x'_0)^l_{(m)}]]
\end{eqnarray}
where $f$ and $g$ are functions of time:
\begin{equation}
f^{ijkl}_{(n)(m)}(t) = \frac{1}{8 \pi^2 t} \int_{\mid {\bf k} \mid <
\Lambda}
\frac{d^3k}{{\bf k}^2} (1 - \frac{\sin \mid {\bf k} \mid t}
{\mid {\bf k} \mid t}) \cos {\bf k}({\bf a_n}-{\bf a_m}) T^{ijkl}({\bf
k})
\end{equation}
\begin{equation}
g^{ijkl}_{(n)(m)}(t) = \frac{1}{8 \pi^2 t} \int_{\mid {\bf k} \mid <
\Lambda} \frac{d^3k}{{\bf k}^2} \frac{1 - \cos \mid {\bf k} \mid
t}{\mid {\bf k} \mid t}
 \cos {\bf k}({\bf a_n}-{\bf a_m}) T^{ijkl}({\bf k})
\end{equation}
and in particular:
\begin{equation}
f^{ijkl}_{(n)(n)}(t) = \frac{1}{15 \pi t} \delta^{ijkl}
\int_0^{\Lambda} dk (1 - \frac{\sin kt}{kt})
\end{equation}
\begin{equation}
g^{ijkl}_{(n)(n)}(t)= \frac{1}{15 \pi t} \delta^{ijkl}
\int_0^{\Lambda} dk \frac{1-\cos kt}{kt}
\end{equation}
\par
The standard prescription for the derivation of the master equation
from the reduced density master propagator consists of taking its time
derivative and using identities relating ${\bf x_0}$ and ${\bf x'_0}$ with the
action of derivatives with respect to ${\bf x_f}$ and ${\bf
x'_f}$. For the interested reader, we
list the relevant identities in the appendix.
\par
After some  calculations, the master equation turns out to be
(inserting back $\hbar$,$m$ and $c$):
\begin{eqnarray}
\frac{\partial}{\partial t} \rho = \sum_n \frac{i \hbar}{2 m}( 1 -
\delta m_{(n)}(t)) ( \frac{\partial^2}{\partial {\bf x_{(n)}}^2} -
\frac{\partial^2}{\partial {\bf x'_{(n)}}^2}) \rho \\ \nonumber
- \frac{i \hbar^4}{4 m^2}  \sum_{n,m} \alpha^{ijkl}_{(n)(m)}(t)
(\frac{\partial^4}{\partial
x^i_{(n)} \partial x^j_{(n)} \partial x^k_{(m)} \partial x^l_{(m)}}-
\frac{\partial^4}{\partial
x'^i_{(n)} \partial x'^j_{(n)} \partial x'^k_{(m)} \partial
x'^l_{(m)}}) \rho \\ \nonumber
- \frac{\hbar^4}{4 m^2} \sum_{n,m} \beta^{ijkl}_{(n)(m)}(t) (
\frac{\partial^4}{\partial
x^i_{(n)} \partial x^j_{(n)} \partial x^k_{(m)} \partial x^l_{(m)}}+
\frac{\partial^4}{\partial
x'^i_{(n)} \partial x'^j_{(n)} \partial x'^k_{(m)} \partial
x'^l_{(m)}} \\ \nonumber
- 2 \frac{\partial^4}{\partial
x^i_{(n)} \partial x^j_{(n)} \partial x'^k_{(m)} \partial
x'^l_{(m)}}) \rho
\end{eqnarray}
\par
This is the main result of this paper: the master equation for N
non-relativistic particles interacting through linearized gravity. The
gravitational field induces a renormalization in the mass of the
particles, modifies the dynamics so that they become dissipative and
is responsible for noise. These three effects are contained in the
functions $\delta m(t)$, $\alpha (t)$ and $\beta(t)$
respectively:
\begin{equation}
\delta m_{(n)}(t) = \frac{4G \hbar}{c^5} g^{ijkl}_{(n)(n)} \delta_{ij}
\delta_{kl}
\end{equation}
\begin{equation}
\alpha^{ijkl}_{(n)(m)}(t) = \frac{4 G}{\hbar c^5}
\dot{f}^{ijkl}_{(n)(m)} t^2
\end{equation}
\begin{equation}
\beta^{ijkl}_{(n)(m)} = \frac{4 G}{\hbar c^5} (t g + \frac{1}{2}
\dot{g}t^2)^{ijkl}_{(n)(m)}
\end{equation}

\section{One particle}
\par
An interesting case is that of a single particle. Since then the
functions in
the master equation are totally symmetric in the
spatial indices, we can, without loss of generality, consider it
constrained to move in only one dimension.
The master equation reads then in operator form
\begin{equation}
\frac{\partial}{\partial t} \rho = - \frac{i}{\hbar} [H_R,\rho] - i
\alpha(t)[H_R^2,\rho] - \beta(t) [H_R,[H_R,\rho]]
\end{equation}
and depends explicitly only on the renormalized Hamiltonian $H_R$.
We can verify
that this form of master equation (in particular the noise part)
is particular to the free
particle case.  For an harmonic oscillator we
would get  an extra dissipation and diffusion
term due to the coupling of the particle's position to the graviton
oscillator Hamiltonian, and of form similar to the one derived in
\cite{HPZ} for quadratic coupling to position.
\par
The diffusion coefficient $\beta(t)$ exhibits a ``jolt'' for times of
the order of $\Lambda^{-1}$. In quantum Brownian models this is a
cause of rapid decoherence of the density matrix of the particle, and
diagonalization in a basis determined by the coupling to the
environment. Our particular form of the diffusion terms tempts us to
propose that it should lead to diagonalization of the particle's
density matrix in the
energy eigenstate basis. But we have to take into account, that the coupling
is extremely weak and that after the jolt the diffusion coefficient
falls to zero, quite slowly actually since it goes at most like
$1/t$.
\par
We can give an estimation of the decoherence in the energy by
approximating $\beta(t)$ with a constant of the order of $\frac{G
\Lambda^2}{\hbar c}$ for times of the order of $\Lambda^{-1}$ and
zero afterwards. We borrow some ideas from the quantum state diffusion
picture of quantum mechanics \cite{GiPe,Pe,HaZo}.
At the times that $\beta(t)$ is constant,
we have a unique unravelling of the density matrix into states
evolving stochastically in a Hilbert space. It is straightforward to
show \cite{GiPe} that an initial wavepacket with energy
spread $\Delta E_0$ will emerge after the jolt with spread $\Delta E$
given by:
\begin{equation}
\frac{1}{(\Delta E)^2} - \frac{1}{(\Delta E_0)^2} \sim \frac{G
\Lambda}{\hbar c^5}
\end{equation}
For a single particle of mass $m$ a good upper bound on $\hbar \Lambda$ is $
\frac{G m^3
c}{\hbar}$ : the classical gravitational self energy of a mass distribution
localized within the Compton wavelength of the particle. This means
that:
\begin{equation}
\frac{1}{(\Delta E)^2} - \frac{1}{(\Delta E_0)^2} \sim \frac{G^2
m^3}{\hbar^3 c^4}
\end{equation}
This is an extremely small quantity, when considering microscopic
particles (even in atomic scales). On the other hand, for macroscopic
and even mesoscopic
particles  the right hand side is quite large and we expect a
loclalization of the particle in its energy eigenstates. For instance
a particle with mass $m = 10^{-8} gr $ and irrespectively of its
initial configuration, will emerge after $ 10 ^{-30} s$ localized in
an energy eigenstate with spread of the order of $ 0.1 MeV $, which is
a tiny portion of its kinetic energy.
But in this case,
the gravity induced decoherence is in general, hidden beneath the
effects of other types of environment \cite{JoZe}. In any case, this
result is in good agreement with the assumptions of the
gravitationally inducced collapse models.
\par
These features were , more or less expected, since gravity couples
very weakly and its strength increases with the mass of the
interacting bodies.
 Still,there was
the possibility, that a persistent noise source might induce
decoherence even in microscopic systems,despite the weakness of the
coupling. Note, that our analysis based on the linearized
approximation, does not rule out the possibility that highly
non-linear Planck scale processes \cite{Pen,Zeh} might be a source of
noise, giving rise to decoherence at smalles mass scales.
\par
The dissipation function $\alpha(t)$ approaches asymptotically a
constant of the order of $\frac{G \Lambda}{\hbar c^5}$. The overall
picture we get, is that of a particle continuously dissipating energy
and suffering at early times noise from the environment until it
becomes correlated with the gravitational field.

\section{Conclusions}
\par
We have studied the quantum theory of N non-relativistic particles,
coupled to the linearized gravitational field using the influence
functional formalism. Our main result was the
master equation (3.9) containing information  of non-local interaction
between the particles. We should note that the gravitational field,
being coupled quadratically to the velocities gives a rather unusual
expression for the influence functional. This results in a master
equation, where both dissipation and diffusion are determined uniquely
by the Hamiltonian operator. This is in accordance with our intuitive
feeling, that the gravitational field acts like continuously
``measuring'' a particle's energy.
\par
One of our motivations for this work, was to establish whether we can
consider the gravitational field as a source of fundamental decoherence
in quantum mechanics. The answer comes out negative for microscopic
systems, but systems with large mass seem to decohere within a fast
rate, in the energy
eigenstate basis.  In addition, it might be interesting to examine,
the evolution
of a single particle under the action of a particular matter
distribution. The formalism we used can be extended with slight
modifications to cover this case. We can, for instance, consider
almost stationary cosmic dust and even a cosmological spacetime. The
collective effect of matter plus gravity might give a strongest
decoherence to the particle.
\par
In addition, it would be of interest to study the response of a
system of detectors to different initial conditions for the graviton
field. The case where a number of modes is excited, seems very
interesting. The information of the state of the field should be
encoded in the correlation kernels of the particles, from the time
evolution of which we would be able to determine the presence of the
graviton fields. This might give a nice toy model for detectors of
gravitational waves.

\section{Aknowledgements}
\par
I would like to thank J.\ J.\ Halliwell and A.\ Zoupas for useful
discussions and suggestions.
\par
The research was supported by the Greek State Scholarship Foundation.

\section{Appendix}
\par
We give here the identities that enable us to compute perturbatively
the master equation. We give the form for the case of one dimension
and one particle. The generalization is straightforward.
\begin{eqnarray}
(x_f - x_0)^2 J(x_f,x'_f,t \mid x_0,x'_0,0) = \hspace{5cm}\\ \nonumber
[-t^2 (1 - 2i G (9 f - 11
i g) (\frac{\partial^2}{\partial x_f^2}
- \frac{i}{t})
-2 G t^2
(\frac{\partial^2}{\partial x_f^2} + \frac{i}{t}) - 2 G t ( 3f -4 i g)
\\ \nonumber
+ \frac{4 G}{t} (f - ig) (x_f - x_0)^4 + \frac{4 i G}{t} g (x_f - x_0)^2
(x'_f -x'_0)^2] J(x_f,x'_f,t \mid x_0,x'_0,0) \\ \nonumber
+ {\cal O}(G^2)
\end{eqnarray}
\begin{eqnarray}
(x_f - x_0)^4 J(x_f,x'_f,t \mid x_0,x'_0,0) = \\ \nonumber
[t^4
\frac{\partial^4}{\partial x_f^4} - 6 i t^3 \frac{\partial^2}{\partial
x_f^2} - 3 t^2] J(x_f,x'_f,t \mid x_0,x'_0,0 + {\cal O}(G)
\end{eqnarray}
\begin{eqnarray}
(x_f - x_0)^2 (x'_f - x'_0)^2 J(x_f,x'_f,t \mid x_0,x'_0,0) =
\\ \nonumber
[ t^4 \frac{\partial^4}{\partial x_f^2 \partial x'_f{}^2} + i t^3
\frac{\partial^2}{\partial x_f^2}- i t^3 \frac{\partial^2}{\partial
x'_f{}^2} + t^2] J(x_f,x'_f,t \mid x_0,x'_0,0) + {\cal O}(G)
\end{eqnarray}
We should keep in mind that eventually,  we keep terms to first order
in $G$. The expressions for the primed quantities are obtained by
permutation of primed with unprimed ones and complex conjugation.
\end{document}